**NUCLEI, PARTICLES,
AND THEIR INTERACTION**

# Electromagnetic Form Factors of a Massive Neutrino

**M. S. Dvornikov\* and A. I. Studenikin\*\***
*Moscow State University, Vorob'evy Gory, Moscow, 119992 Russia*
*e-mail: maxim_dvornikov@aport.ru; studenik@srd.sinp.msu.ru*
Received April 14, 2004

**Abstract**—Electromagnetic form factors of a massive neutrino are studied in a slightly extended standard model in an arbitrary $R_\xi$ gauge and taking into account the dependence on the masses of all interacting particles. The contribution from all Feynman diagrams to the charge, magnetic, and anapole form factors, in which the dependence of the masses of all particles as well as on gauge parameters is accounted for exactly, are obtained for the first time in explicit form. The asymptotic behavior of the magnetic form factor for large negative squares of the momentum of an external photon is analyzed and the expression for the anapole moment of a massive neutrino is derived. The results are generalized to the case of mixing between various generations of the neutrino. Explicit expressions are obtained for the charge, magnetic, and electric dipole and anapole transient form factors as well as for the transient electric dipole moment. © *2004 MAIK "Nauka/Interperiodica"*.

## 1. INTRODUCTION

Analysis of electromagnetic properties of neutrinos is of considerable interest in the light of the recent experimental verification of the existence of a nonzero neutrino mass and mixing between different neutrino flavors [1–3]. At the present time, it remains not completely clear whether the neutrino is a Dirac or Majorana particle. It should be noted that these two types of elementary particles possess basically different electromagnetic characteristics [4]. It is well known that a fermion with a spin of 1/2 can possess no more than four electromagnetic form factors. As a rule, these quantities are defined in terms of the charge, magnetic dipole, electric dipole, and anapole form factors (see also Section 2). However, a Majorana neutrino can exhibit its electromagnetic properties only in terms of the interaction of the anapole form factor with an external electromagnetic field.

The calculation of radiative correction to static characteristics of an elementary particle, viz., its charge determined by the values of the corresponding form factors for zero momentum transfer, is of considerable interest. In this connection, the publications [5–10] in which the electromagnetic moments of neutrinos were calculated in various theoretical models are worth mentioning. In the series of recent publications [11–13], the electric charge and the magnetic moment of a neutrino in an arbitrary $R_\xi$ gauge were studied. It should be recalled that the corresponding form factors for zero momentum transfer are elements of the scattering matrix and, hence, can be measured in experiment. Thus, the electric charge and magnetic moment should be independent of the choice of the gauge. This was demonstrated in [11–13] even for a massive neutrino [13]. Analysis of radiative corrections to electromagnetic parameters of the neutrino may directly indicate which physical theory should be used beyond the range of the standard model and provide important information on the parameters and structure of the proposed model of interaction between elementary particles. For example, for particles described in the framework of the theories with broken CP invariance, an electric dipole moment inevitably appears.

For nonzero momentum transfer, the electromagnetic form factors are not invariants of the gauge transformation group and, hence, are not measurable quantities. However, analyzing some processes (e.g., calculating higher order corrections) the values of the electromagnetic form factors of neutrinos with a nonzero momentum transfer must be taken into account. One of such processes corresponding to radiative corrections to scattering of a neutrino by a lepton is shown in Fig. 1.

The charge and magnetic form factors of elementary particles were analyzed in [6, 14] in the frameworks of various gauge theories The case of zero-mass neutrino was studied and not only static electromagnetic parameters, but also asymptotic behavior of the magnetic form factors for high negative squares of the external photon momentum were considered. It is well known that the magnetic moment of a neutrino in the slightly extended standard model is proportional to the particle mass. In a number of theoretical models (see, for example, [15]), the magnitude of the magnetic moment depends on the neutrino mass only slightly and is completely determined by the mass of the heavy particle in the polarization loop.

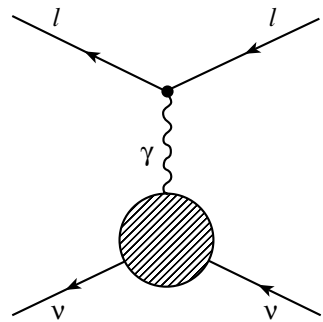

**Fig. 1.** A diagram illustrating the contribution to the elastic scattering of a neutrino by a lepton. Hatched circle schematically represents the neutrino electromagnetic vertex function.

One of electromagnetic properties of an elementary particle is its charge radius which is studied in many publications. The expression for the charge radius of a zero-mass neutrino in the standard model was derived in [16]. It was found that the charge radius for a zero-mass particle is a diverging quantity; moreover, it depends on the choice of gauge. In this connection, the concept of electroweak radius was introduced in [16, 17]. This quantity is determined by radiative corrections to the process of scattering of a lepton by a neutrino. The electroweak radius of a neutrino is connected with the effective Weinberg angle. Obviously, the quantity defined in this way is finite and gauge invariant. The calculation of the electroweak radius of a zero-mass neutrino in the framework of the standard model in an arbitrary $R_\xi$ gauge was carried out in [17].

The neutrino anapole moment was considered by many authors. Among the corresponding publications, the work [18] in which the anapole moment is shown to be gauge-dependent in the frame of the standard model and, hence,s unobservable, is worth mentioning. In a series of publications [19, 20] using the dispersion relation method, the expression for the neutrino anapole moment was derived in the 't Hoft–Feynman gauge and the dependence of the anapole form factor on the square of the external photon momentum was studied. We must also mention the article [21] in which the expression for the anapole moment of a zero-mass neutrino was derived on the basis of the electroweak radius. As a matter of fact, a zero-mass particle is characterized by a certain relation between these quantities (see also Section 5.1). Thus, the knowledge of one of the electromagnetic parameters (in the present case, electroweak radius) makes it possible to easily reconstruct another characteristic and to derive an expression for the neutrino electroweak anapole moment.

In this study, we analyze the electromagnetic vertex form factors of a massive Dirac neutrino in the framework of the slightly extended standard model supplemented with a SU(2)-singlet right neutrino. All calculations are made in an arbitrary $R_\xi$ gauge, while enables us to study the dependence of the obtained results on gauge parameters for both W and Z bosons. It should be noted that the masses of the neutrino and the charged lepton were never fixed; consequently, our analysis makes it possible to consider the limit of not only light, but also heavy neutrinos. In Section 2, the Feynman amplitudes are determined for all contributions to the electromagnetic vertex function of a massive neutrino. The expressions for the contributions of self-energy $\gamma$–$Z$ diagrams are calculated in explicit form containing a single definite integral with respect to the Feynman parameter. The ultraviolet divergences emerging in the analysis of the neutrino electromagnetic vertex function are also treated in Section 2. In Sections 3 and 4, the contributions of all Feynman diagrams to the charge and magnetic form factors of a massive neutrino are studied. The asymptotic behavior of the magnetic form factor is analyzed for large negative squares of the external photon momentum. The anapole form factor and the anapole moment of a massive neutrino are considered in Section 5. The results of this study can be generalized to the case of mixing between different generations of the neutrino. In particular, transient electromagnetic form factors are studied in Section 6 in the framework of the slightly extended standard model permitting mixing between different generations of charged leptons and neutrino. Explicit expressions are derived for the charge, magnetic, and electric dipole and transient anapole form factors. The cases of mass-degenerate and nondegenerate neutrino states are studied. Moreover, the expression for the transient electric dipole moment is also derived in Section 6.

## 2. NEUTRINO VERTEX FUNCTION

The matrix element of electromagnetic current averaged over the neutrino states can be written in the form

$$\langle \nu(p')|J_\mu^{EM}|\nu(p)\rangle = \bar{u}(p')\Lambda_\mu(q)u(p), \qquad (2.1)$$

where the most general expression for the electromagnetic vertex function $\Lambda_\mu(q)$ is

$$\Lambda_\mu(q) = f_Q(q^2)\gamma_\mu + f_M(q^2)i\sigma_{\mu\nu}q^\nu \\ - f_E(q^2)\sigma_{\mu\nu}q^\nu\gamma_5 + f_A(q^2)(q^2\gamma_\mu - q_\mu\slashed{q})\gamma_5. \qquad (2.2)$$

Here, $f_Q(q^2)$, $f_M(q^2)$, $f_E(q^2)$, and $f_A(q^2)$ are the charge, magnetic and electric dipole, and anapole form factors of the neutrino; $q_\mu = p'_\mu - p_\mu$, $\sigma_{\mu\nu} = (i/2)[\gamma_\mu, \gamma_\nu]$, and $\gamma_5 = -i\gamma^0\gamma^1\gamma^2\gamma^3$. The values of these form factors for $q^2 = 0$ determine the static electromagnetic properties of neutrinos. In the case of a Dirac neutrino, which will be considered here, the assumptions concerning the CP invariance and the hermiticity of the electromagnetic current operator $J_\mu^{EM}$ lead to zero value of the dipole electric form factor. At zero momentum transfer, only $f_Q(0)$ and $f_M(0)$, which are known as the electric charge and magnetic moment, give a contribution to Hamiltonian $H_{\text{int}} \sim J_\mu^{EM}A^\mu$ describing the interaction of the neutrino with the external electric field $A^\mu$.

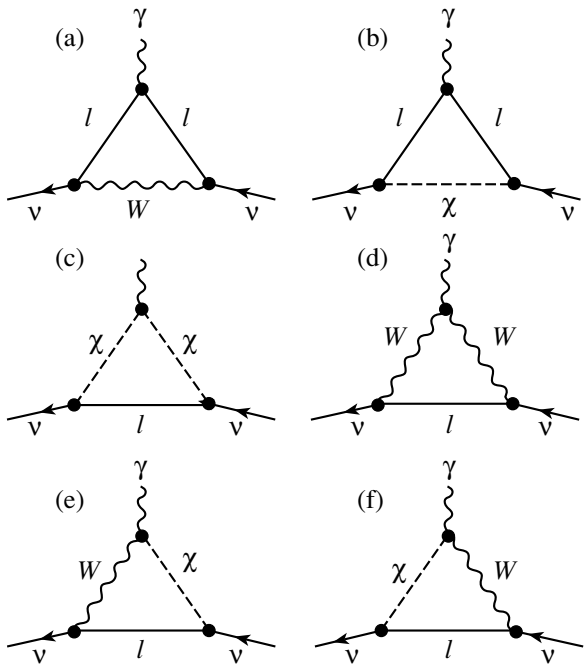

**Fig. 2.** Triangular diagrams.

The expressions for the electromagnetic function of a massive and a zero-mass neutrino differ radically. If we consider a massless elementary particle, relation (2.2) implies that the matrix element of the electromagnetic current can be written using only one form factor (see, for example, [21]),

$$\bar{u}(p')\Lambda_\mu(q)u(p) = f_D(q^2)\bar{u}(p')\gamma_\mu(1+\gamma_5)u(p).$$

It follows hence that the charge and anapole form factors are connected with function $f_D(q^2)$ via obvious relations:

$$f_Q(q^2) = f_D(q^2), \quad f_A(q^2) = f_D(q^2)/q^2.$$

However, in the case of a massive elementary particle, a simple relation connecting the charge and anapole form factor does not exist since we cannot disregard matrix terms of the form $q_\mu \slashed{q}\gamma_5$ in the term proportional to the anapole form factor. Moreover, direct calculation of the neutrino electromagnetic form factors revealed that, in addition to the well known form factors given in relation (2.2), each Feynman diagram gives a nonzero contribution to the additional term proportional to matrix $\gamma_\mu\gamma_5$. These contributions differ from zero even for $q_2 = 0$. It was found in our earlier publication [13] that the sum of these contributions from all Feynman diagrams to the additional "form factor" is zero for $q^2 = 0$. The equality to zero of the "form factor" in question for $q^2 \neq 0$ in a special gauge was also proved in [13].

We will consider the calculation of one-loop Feynman diagrams for the vertex electromagnetic function of a massive neutrino in the framework of the slightly extended standard model with an SU(2)-singlet right neutrino in an arbitrary $R_\xi$ gauge. These diagram can be divided into two types: triangular (Fig. 2) and $\gamma$–Z diagrams (Fig. 4). Using Feynman's rules formulated in [22], we can determine the contributions to the neutrino vertex function $\Lambda_\mu(q)$. Applying size regularization in the corresponding Feynman integrals, we find that the contributions from the triangular diagrams (Fig. 2) can be written in the form

$$\Lambda_\mu^{(1)} = i\frac{eg^2}{2}\int\frac{d^N k}{(2\pi)^N}\left[g^{\kappa\lambda}-(1-\alpha)\frac{k^\kappa k^\lambda}{k^2-\alpha M_W^2}\right]\frac{\gamma_\kappa^L(\slashed{p}'-\slashed{k}+m_l)\gamma_\mu(\slashed{p}-\slashed{k}+m_l)\gamma_\lambda^L}{[(p'-k)^2-m_l^2][(p-k)^2-m_l^2][k^2-M_W^2]}, \quad (2.3)$$

$$\Lambda_\mu^{(2)} = i\frac{eg^2}{2M_W^2}\int\frac{d^N k}{(2\pi)^N}\frac{(m_\nu P_L - m_l P_R)(\slashed{p}'-\slashed{k}+m_l)\gamma_\mu(\slashed{p}-\slashed{k}+m_l)(m_l P_L - m_\nu P_R)}{[(p'-k)^2-m_l^2][(p-k)^2-m_l^2][k^2-\alpha M_W^2]}, \quad (2.4)$$

$$\Lambda_\mu^{(3)} = i\frac{eg^2}{2M_W}\int\frac{d^N k}{(2\pi)^N}(2k-p-p')_\mu\frac{(m_\nu P_L - m_l P_R)(\slashed{k}+m_l)(m_l P_L - m_\nu P_R)}{[(p'-k)^2-\alpha M_W^2][(p-k)^2-\alpha M_W^2][k^2-m_l^2]}, \quad (2.5)$$

$$\Lambda_\mu^{(4)} = i\frac{eg^2}{2}\int\frac{d^N k}{(2\pi)^N}\gamma_\kappa^L(\slashed{k}+m_l)\gamma_\lambda^L\left[\delta_\beta^\kappa-(1-\alpha)\frac{(p'-k)^\kappa(p'-k)_\beta}{(p'-k)^2-\alpha M_W^2}\right]\left[\delta_\gamma^\lambda-(1-\alpha)\frac{(p-k)^\lambda(p-k)_\gamma}{(p-k)^2-\alpha M_W^2}\right]$$

$$\times\frac{\delta_\mu^\beta(2p'-p-k)^\gamma + g^{\beta\gamma}(2k-p-p')_\mu + \delta_\mu^\gamma(2p-p'-k)^\beta}{[(p'-k)^2-M_W^2][(p-k)^2-M_W^2][k^2-m_l^2]}, \quad (2.6)$$

$$\Lambda_\mu^{(5)+(6)} = i\frac{eg^2}{2}\int\frac{d^Nk}{(2\pi)^N}\left\{\frac{\gamma_\beta^L(\not{k}-m_l)(m_lP_L-m_\nu P_R)}{[(p'-k)^2-M_W^2][(p-k)^2-\alpha M_W^2][k^2-m_l^2]}\left[\delta_\mu^\beta-(1-\alpha)\frac{(p'-k)^\beta(p'-k)_\mu}{(p'-k)^2-\alpha M_W^2}\right]\right.$$

$$\left.-\frac{(m_\nu P_L-m_l P_R)(\not{k}-m_l)\gamma_\beta^L}{[(p'-k)^2-\alpha M_W^2][(p-k)^2-M_W^2][k^2-m_l^2]}\left[\delta_\mu^2-(1-\alpha)\frac{(p-k)^\beta(p-k)_\mu}{(p-k)^2-\alpha M_W^2}\right]\right\}. \quad (2.7)$$

Here, $m_\nu$, $M_W$, and $m_l$ are the masses of the neutrino, the $W$ boson, and the charged lepton (which is the lower component of the isodoublet relative to the neutrino); $e$ is the proton charge; $g$ is the coupling constant in the standard model; $\theta_W$ is the Weinberg angle; $\alpha = 1/\xi$ is the gauge parameter for the $W$ boson; and $P_{L,R} = (1\pm\gamma_5)/2$ are the projection operators.

The contributions from the $\gamma$–$Z$ diagrams (Fig. 4) to the vertex function $\Lambda_\mu(q)$ are shown in Fig. 3 and are given by the following expressions:

$$\Lambda_\mu^{(j)}(q) = \frac{g}{2\cos\theta_W}\Pi_{\mu\nu}^{(j)}(q)\frac{1}{q^2-M_Z^2}$$

$$\times\left\{g^{\nu\alpha}-(1-\alpha_Z)\frac{q^\nu q^\alpha}{q^2-\alpha_Z M_Z^2}\right\}\gamma_\alpha^L, \quad (2.8)$$

$$j = 7, \ldots, 14,$$

where

$$\Pi_{\mu\nu}^{(7)}(q) = -ieg\cos\theta_W$$

$$\times\int\frac{d^Nk}{(2\pi)^N}\frac{1}{[(k-q)^2-M_W^2][k^2-M_W^2]}$$

$$\times\left[g_{\gamma\alpha}-(1-\alpha)\frac{(k-q)_\gamma(k-q)_\alpha}{(k-q)^2-\alpha M_W^2}\right]$$

$$\times\left[g_{\beta\lambda}-(1-\alpha)\frac{k_\beta k_\lambda}{k^2-\alpha M_W^2}\right] \quad (2.9)$$

$$\times[(k+q)^\gamma\delta_\mu^\beta+(q-2k)_\mu g^{\beta\gamma}+(k-2q)^\beta\delta_\mu^\gamma]$$

$$\times[(k+q)^\alpha\delta_\nu^\lambda+(q-2k)_\nu g^{\alpha\lambda}+(k-2q)^\lambda\delta_\nu^\alpha],$$

$$\Pi_{\mu\nu}^{(8)}(q) = -2ieg\frac{\sin^2\theta_W}{\cos\theta_W}M_W^2$$

$$\times\int\frac{d^Nk}{(2\pi)^N}\frac{1}{[(k-q)^2-\alpha M_W^2][k^2-M_W^2]} \quad (2.10)$$

$$\times\left[g_{\mu\nu}-(1-\alpha)\frac{k_\mu k_\nu}{k^2-\alpha M_W^2}\right],$$

$$\Pi_{\mu\nu}^{(9)}(q) = ieg\frac{\cos^2\theta_W-\sin^2\theta_W}{\cos\theta_W}$$

$$\times\int\frac{d^Nk}{(2\pi)^N}\frac{g_{\mu\nu}}{k^2-\alpha M_W^2}, \quad (2.11)$$

$$\Pi_{\mu\nu}^{(10)}(q) = -ieg\cos\theta_W$$

$$\times\int\frac{d^Nk}{(2\pi)^N}\frac{\delta_\mu^\alpha\delta_\nu^\beta+\delta_\mu^\beta\delta_\nu^\alpha-2g^{\alpha\beta}g_{\mu\nu}}{k^2-M_W^2} \quad (2.12)$$

$$\times\left[g_{\alpha\beta}-(1-\alpha)\frac{k_\alpha k_\beta}{k^2-\alpha M_W^2}\right],$$

$$\Pi_{\mu\nu}^{(11)+(12)}(q) = 2ieg\cos\theta_W$$

$$\times\int\frac{d^Nk}{(2\pi)^N}\frac{k_\mu(k-q)_\nu}{[(k-q)^2-\alpha M_W^2][k^2-\alpha M_W^2]}, \quad (2.13)$$

$$\Pi_{\mu\nu}^{(13)}(q) = ieg\frac{\sin^2\theta_W-\cos^2\theta_W}{2\cos\theta_W}\int\frac{d^Nk}{(2\pi)^N}(2k-q)_\mu$$

$$\times(2k-q)_\nu\frac{1}{[(k-q)^2-\alpha M_W^2][k^2-\alpha M_W^2]}, \quad (2.14)$$

$$\Pi_{\mu\nu}^{(14)}(q) = \frac{ieg}{2\cos\theta_W}\sum_f Q_f\int\frac{d^Nk}{(2\pi)^N}$$

$$\times\frac{1}{[(k-q)^2-m_f^2][k^2-m_f^2]}\text{Tr}\left[\gamma_\mu(\not{k}+m_f)\gamma_\nu \quad (2.15)\right.$$

$$\left.\times\left(\pm\frac{1}{2}-2Q_f\sin^2\theta_W\pm\frac{1}{2}\gamma_5\right)(\not{k}-\not{q}+m_f)\right].$$

Here, $M_Z$ and $\alpha_Z$ denote the mass and the gauge parameter of the $Z$ boson. The minus and plus signs in expression (2.15) correspond to the "upper" ($u$, $c$, and $t$ quarks) and "lower" (electron, muon, $\tau$ lepton as well as $d$, $s$, and $b$ quarks) components of the isodoublet and $m_f$ and $Q_f$ are the mass and the electric charge (in units of $e$) of a fermion in the loop.

It will be convenient in the subsequent analysis to expand each contribution from the $\gamma$–$Z$ diagrams for an

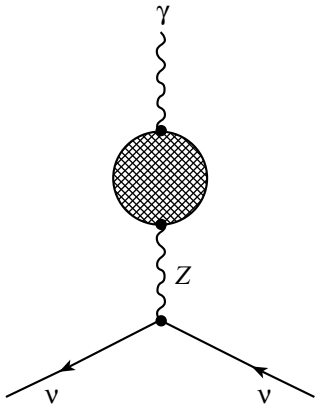

**Fig. 3.** Contributions of the $\gamma$–$Z$ diagrams to the neutrino electromagnetic vertex function. Hatched circle schematically represents the function $\Pi^{(j)}_{\mu\nu}(q)$.

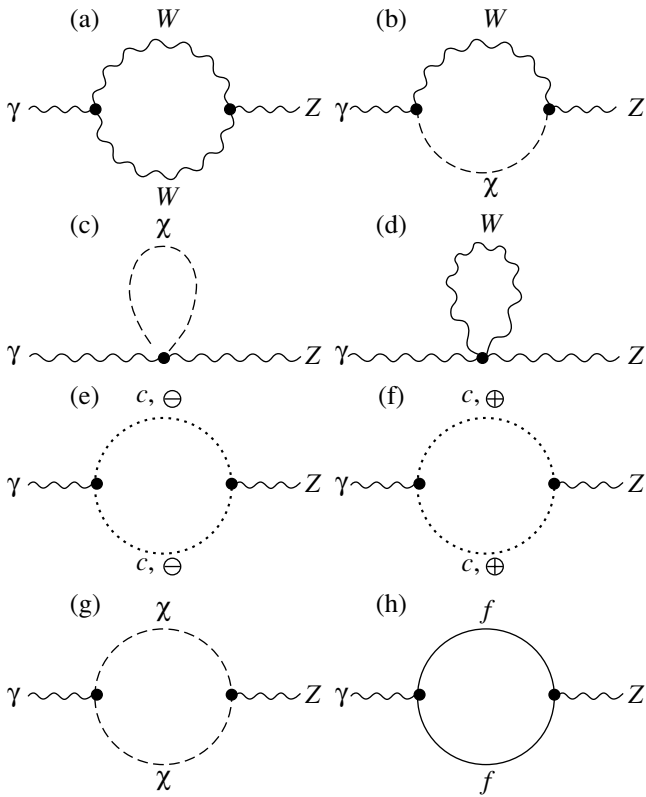

**Fig. 4.** $\gamma$–$Z$ diagrams: $f$ denotes an electron, muon, $\tau$ lepton as well as the $u$, $c$, $t$, $d$, $s$, and $b$ quarks.

arbitrary $q^2$ and to explicitly separate the transverse part:

$$\Pi^{(j)}_{\mu\nu}(q) = A^{(j)}(\alpha, q^2)\left(g_{\mu\nu} - \frac{q_\nu q_\nu}{q^2}\right) + B^{(j)}(\alpha, q^2)g_{\mu\nu}, \quad (2.16)$$
$$j = 1, \ldots, 14.$$

Using expressions (2.9)–(2.15) for the contributions from the $\gamma$–$Z$ diagrams in the form of Feynman integrals as well as formula (2.16), we can write the functions $A^{(j)}(a, q^2)$ and $B^{(j)}(a, q^2)$, where $j = 7, \ldots, 14$, in explicit form:

$$A^{(7)}(\alpha, q^2) = 2M_W^2 \cos^3\theta_W \sin\theta_W M_Z^2 \tilde{G}_F \tau$$

$$\times\left[\omega\left(-\frac{14}{3} + \alpha\right) + \frac{1}{6} + \frac{\alpha}{2}\right.$$

$$-2\tau\int_0^1 dx(1-x^2)^2\{\ln(1-\zeta-x(1-\alpha)) - \ln(1-\zeta)\}$$

$$+ 2\int_0^1 dx(5x^2 - 5x - 1)\ln(1-\zeta) \quad (2.17)$$

$$-2\int_0^1 dx(4x^2 - 3)\{(1-\zeta-x(1-\alpha))$$

$$\times\ln(1-\zeta-x(1-\alpha)) - (1-\zeta)\ln(1-\zeta)\}$$

$$+ \frac{\tau}{2}\int_0^1 dx\{2(1-\zeta-x(1-\alpha))\ln(1-\zeta-x(1-\alpha))$$

$$\left. - (1-\zeta)\ln(1-\zeta) - (\alpha-\zeta)\ln(\alpha-\zeta)\}\right],$$

$$A^{(8)}(\alpha, q^2) = -4M_W^2 \cos\theta_W \sin^3\theta_W M_Z^2 \tilde{G}_F \tau$$
$$\times \int_0^1 dx x^2\{\ln(1-\zeta-x(1-\alpha)) - \ln(\alpha-\zeta)\}, \quad (2.18)$$

$$A^{(9)}(\alpha, q^2) = 0, \quad (2.19)$$

$$A^{(10)}(\alpha, q^2) = 0, \quad (2.20)$$

$$A^{(11)+(12)}(\alpha, q^2) = 2M_W^2 \cos^3\theta_W \sin\theta_W M_Z^2 \tilde{G}_F \tau$$
$$\times\left[\frac{\omega}{3} + 2\int_0^1 dx x(1-x)\ln(\alpha-\zeta)\right], \quad (2.21)$$

$$A^{(13)}(\alpha, q^2) = M_W^2(\sin^2\theta_W - \cos^2\theta_W)\cos\theta_W \sin\theta_W$$
$$\times M_Z^2 \tilde{G}_F \tau\left[-\frac{\omega}{3} - \int_0^1 dx(2x-1)^2 \ln(\alpha-\zeta)\right], \quad (2.22)$$

$$A^{(14)}(\alpha, q^2) = 8M_W^2 \cos\theta_W \sin\theta_W M_Z^2 \tilde{G}_F \tau$$

$$\times \left[ \frac{\omega}{6}\left(-3 - \frac{28}{3}\sin^2\theta_W\right) \right. \quad (2.23)$$

$$\left. + \sum_f Q_f\left(\pm\frac{1}{2} - 2Q_f\sin^2\theta_W\right)\left\{\frac{1}{6}\ln\left(\frac{m_f}{M}\right)^2 \right.\right.$$

$$\left.\left. + \int_0^1 dx\, x(1-x)\ln(1-(M/m_f)^2\zeta)\right\}\right],$$

$$B^{(7)}(\alpha, q^2) = 2M_W^2\cos^3\theta_W\sin\theta_W M_Z^2 \tilde{G}_F$$

$$\times \left[\omega\left(\frac{\tau}{2} - \frac{12 + 3\alpha(1+\alpha)}{2}\right)\right.$$

$$+ \frac{3}{4}(2 + \alpha(1+\alpha)) - \frac{\tau}{24}(25 + 3\alpha)$$

$$- 3\tau\int_0^1 dx(2x-1)^2\ln(1-\zeta) - 9\int_0^1 dx(1-\zeta)\ln(1-\zeta) \quad (2.24)$$

$$- 3\tau\int_0^1 dx\, x^2\{(1-\zeta-x(1-\alpha))\ln(1-\zeta-x(1-\alpha))$$

$$- (1-\zeta)\ln(1-\zeta)\} - \frac{9}{2}\int_0^1 dx\{(1-\zeta-x(1-\alpha))^2$$

$$\left.\times \ln(1-\zeta-x(1-\alpha)) - (1-\zeta)^2\ln(1-\zeta)\}\right],$$

$$B^{(8)}(\alpha, q^2) = 2M_W^2\cos\theta_W\sin^3\theta_W M_Z^2 \tilde{G}_F$$

$$\times \left[-\omega\frac{3+\alpha}{2} - \frac{1-\alpha}{2} - 2\int_0^1 dx\ln(1-\zeta-x(1-\alpha))\right.$$

$$+ \int_0^1 dx\{(1-\zeta-x(1-\alpha))\ln(1-\zeta-x(1-\alpha)) \quad (2.25)$$

$$- (\alpha - \zeta)\ln(\alpha - \zeta)\}$$

$$\left.+ 2\tau\int_0^1 dx\, x^2\{\ln(1-\zeta-x(1-\alpha)) - \ln(\alpha-\zeta)\}\right],$$

$$B^{(9)}(\alpha, q^2) = 2M_W^2(\cos^2\theta_W - \sin^2\theta_W)\cos\theta_W \quad (2.26)$$

$$\times \sin\theta_W M_Z^2 \tilde{G}_F[\alpha(\omega-1) + \alpha\ln\alpha],$$

$$B^{(10)}(\alpha, q^2) = 6M_W^2\cos^3\theta_W\sin\theta_W M_Z^2 \tilde{G}_F$$

$$\times \left[\omega\frac{3+\alpha^2}{2} - \frac{1}{4} - \frac{5\alpha^2}{12} + \frac{\alpha^2\ln\alpha}{2}\right], \quad (2.27)$$

$$B^{(11)+(12)}(\alpha, q^2) = 2M_W^2\cos^3\theta_W\sin\theta_W M_Z^2 \tilde{G}_F$$

$$\times \left[\omega\left(\alpha - \frac{\tau}{2}\right) - \alpha + \frac{\tau}{6} + \int_0^1 dx(\alpha-\zeta)\ln(\alpha-\zeta)\right. \quad (2.28)$$

$$\left.- 2\tau\int_0^1 dx\, x(1-x)\ln(\alpha-\zeta)\right],$$

$$B^{(13)}(\alpha, q^2) = 2M_W^2(\sin^2\theta_W - \cos^2\theta_W)\cos\theta_W\sin\theta_W$$

$$\times M_Z^2\tilde{G}_F\left[\alpha(\omega-1) + \frac{\tau}{6} + \int_0^1 dx(\alpha-\zeta)\ln(\alpha-\zeta)\right. \quad (2.29)$$

$$\left.+ \frac{\tau}{2}\int_0^1 dx(2x-1)^2\ln(\alpha-\zeta)\right],$$

$$B^{(14)}(\alpha, q^2) = 0. \quad (2.30)$$

where

$$\tilde{G}_F = \frac{G_F}{4\pi^2\sqrt{2}}, \quad \omega = -\frac{1}{\varepsilon} - \ln(4\pi^2) + \mathbb{C} - \ln\frac{\lambda^2}{M_W^2},$$

$G_F$ is the Fermi constant, $\zeta = \tau x(1-x)$, $\tau = q^2/M_W^2$, $\varepsilon$ and $\lambda$ are the constants introduced during size regularization, and $\mathbb{C}$ is the Euler constant.

While deriving relations (2.17)–(2.30), we used the properties of the algebra of $\gamma$ matrices in the $N$-dimensional space and the expressions for the characteristic Feynman integrals given in [22, 23]. Note that $\varepsilon = 2 - N/2 > 0$, where $N$ is the dimension of space. When regularization is removed, $\varepsilon \longrightarrow 0$.

Let us now consider ultraviolet divergences emerging in the calculation of the electromagnetic vertex function. The sum of the contributions of the diverging parts of triangular diagrams (2.3)–(2.7) to the electromagnetic vertex function of a massive neutrino has the form

$$\Lambda_\mu^{(\text{div.prop.vert.})}(q) = -\frac{eG_F}{4\pi^2\sqrt{2}}M_W^2\omega\frac{3+\alpha}{2}\gamma_\mu^L. \quad (2.31)$$

Note that this expression is independent of the external photon momentum $q_\mu$.

In the subsequent analysis of diverging contributions from the $\gamma$–$Z$ diagrams (2.9)–(2.15), it is convenient to use relations (2.8) and (2.16). Using these formulas, we obtain the following expression for the sum of the contributions from the γ–Z diagrams to the electromagnetic vertex function of a massive neutrino:

$$\Lambda_\mu^{(\gamma-Z)}(q) = \frac{g}{4\cos\theta_W}\left[\frac{A(\alpha, q^2) + B(\alpha, q^2)}{q^2 - M_Z^2}\gamma_\mu\right.$$

$$+ \frac{1}{q^2 - M_Z^2}\left\{\frac{A(\alpha, q^2)}{q^2} + (1 - \alpha_Z)\frac{B(\alpha, q^2)}{q^2 - \alpha_Z M_Z^2}\right\} \quad (2.32)$$

$$\left.\times (q^2\gamma_\mu - q_\mu \slashed{q})\gamma_5 + \frac{\alpha_Z B(\alpha, q^2)}{q^2 - \alpha_Z M_Z^2}\gamma_\mu\gamma_5\right].$$

The diverging parts of functions $A(\alpha, q^2)$ and $B(\alpha, q^2)$ have the form

$$A^{\text{div}}(\alpha, q^2) = 2M_W^2\cos\theta_W\sin\theta_W M_Z^2\tilde{G}_F\tau\omega$$
$$\times\left\{\left(\alpha - \frac{37}{6}\right)\cos^2\theta_W - \frac{151}{18}\sin^2\theta_W\right\}, \quad (2.33)$$

$$B^{\text{div}}(\alpha, q^2) = -2M_W^2\cos\theta_W\sin\theta_W M_Z^2\tilde{G}_F\frac{3+\alpha}{2}. \quad (2.34)$$

Formulas (2.31)–(2.34) imply that all the form factors except the magnetic one contain divergences and depend on the choice of the gauge (both on $\alpha$ and on $\alpha_Z$). In spite of this, we can choose the gauge parameters in such a way that the complete expression for $\Lambda_\mu(q)$ including the contributions from the triangular (Fig. 2) and γ–Z diagrams (Fig. 4) does not contain ultraviolet divergences. Indeed, fixing the gauge parameters as

$$\alpha = \frac{1}{9}(138 + 151\tan^2\theta_W), \quad \alpha_Z = +\infty,$$

we find that all the terms in $\Lambda_\mu(q)$ containing the pole $1/\varepsilon$ mutually cancel out. Thus, in the given gauge, the electromagnetic vertex function of a massive neutrino is finite in the one-loop approximation for an arbitrary external photon momentum $q_\mu$.

An analogous statement can be formulated for the case of the electron electromagnetic vertex function in quantum electrodynamics. The expression for the electron vertex function in the one-loop approximation is given in [23] in an arbitrary gauge. Using formula (24'), p.358, in [23], we find that all form factors in the vertex function are finite for $d_l = 3$, where $d_l$ is the photon gauge parameter.

## 3. CHARGE FORM FACTOR OF A NEUTRINO

In this section, we consider the charge form factor of a massive neutrino. Using the results obtained in the preceding section for various contributions to the neutrino vertex function $\Lambda_\mu(q)$, we single out in formulas (2.3)–(2.15) the coefficients proportional to matrix $\gamma_\mu$, which are the corresponding contributions to the charge form factor $f_Q(q^2)$ in accordance with the expansion given in relation (2.2).

First of all, we consider the contributions of one-loop triangular diagrams (Fig. 2) to the neutrino charge factor. Using the familiar identity

$$\bar{u}(p)(p'_\mu + p_\mu)u(p) = \bar{u}(p')(2m_\nu\gamma_\mu - i\sigma_{\mu\nu}q^\nu)u(p),$$

and integrating over the momenta of virtual particles (the details of this procedure for size regularization are given in [23]), we obtain the exact expressions for the contributions of the diagrams considered here to the charge form factor of a massive neutrino in terms of definite integral:

$$f_Q^{(\text{prop.vert.})}(q^2) = \frac{eG_F}{4\pi^2\sqrt{2}}M_W^2\sum_{i=1}^{6}\bar{f}_Q^2(q^2).$$

Here,

$$\bar{f}_Q^{(1)}(q^2) = \omega\frac{\omega}{2} + 1 + \frac{1-\alpha}{12} + \int_0^1 dz\int_0^z dy \ln\mathfrak{D}_1$$

$$- \int_0^1 dz\int_0^z dy[a + b(1-z)^2 + \tau(1-z+y(z-y))]\frac{1}{\mathfrak{D}_1}$$

$$+ \frac{1}{2}\int_0^1 dz\int_0^z dy[bz^2(1 + b(1-z)^2) + a\tau y(z-y)$$

$$+ b\tau(2zy(z-y)(1-z) + 5y(z-y) - z^2(1-z)) \quad (3.1)$$

$$+ \tau^2 y(z-y)(1-z+yz-y^2)\left[\frac{1}{\mathfrak{D}_1(\alpha)} - \frac{1}{\mathfrak{D}_1}\right]$$

$$- \frac{1}{2}\int_0^1 dz\int_0^z dy[a + b + 6bz(1-z) + \tau(1-3z+6y(z-y))]$$

$$\times [\ln\mathfrak{D}_1(\alpha) + \ln\mathfrak{D}_1],$$

$$\bar{f}_Q^{(2)}(q^2) = \frac{a+b}{2}\left(\frac{\omega}{2} + \frac{1}{2} + \int_0^1 dz\int_0^z dy \ln\mathfrak{D}_1(\alpha)\right)$$

$$- \frac{1}{2}\int_0^1 dz\int_0^z dy(a^2 + abz^2 + b^2z^2 \quad (3.2)$$

$$- 4abz + ab + (a+b)\tau y(z-y))\frac{1}{\mathfrak{D}_1(\alpha)},$$

$$\bar{f}_Q^{(3)}(q^2) = \frac{a+b}{2}\left(-\frac{\omega}{2} - \int_0^1 dz \int_0^z dy \ln \mathcal{D}_2(\alpha)\right)$$

$$+ b\int_0^1 dz \int_0^z dy (3az - az^2 - 2a + bz(1-z))\frac{1}{\mathcal{D}_2(\alpha)}, \quad (3.3)$$

$$\bar{f}_Q^{(4)}(q^2) = -\omega\frac{3}{4}(1+\alpha) - 1 - 3\int_0^1 dz \int_0^z dy \ln \mathcal{D}_2$$

$$+ \int_0^1 dz \int_0^z dy (3bz(1-z) - \tau(z - y(z-y)))\frac{1}{\mathcal{D}_2}$$

$$- \frac{9}{2}\int_0^1 dz \int_0^z dy [(\mathcal{D}_2(\alpha) + y(1-\alpha))$$

$$\times \ln(\mathcal{D}_2(\alpha) + y(1-\alpha)) - \mathcal{D}_2 \ln \mathcal{D}_2]$$

$$- \int_0^1 dz \int_0^z dy [2b^2(1-z)^2(z(1-z) - y)$$

$$- b\tau(y(z-y)(5z - 3z^2 - 3y) + z(1-z)^2 - y(2-y-y^2))$$

$$- \tau^2 y(z-y)(1 - z + yz + y + y^2)]$$

$$\times \left[\frac{1}{\mathcal{D}_2(\alpha) + y(1-\alpha)} - \frac{1}{\mathcal{D}_2}\right] \quad (3.4)$$

$$+ \frac{1}{2}\int_0^1 dz \int_0^z dy [3b(1-z^2) + \tau(4 - (6(z-y) + 11y(z-y)))]$$

$$\times [\ln(\mathcal{D}_2(\alpha) + y(1-\alpha)) - \ln \mathcal{D}_2]$$

$$- \frac{b\tau}{2}\int_0^1 dz \int_0^z dy (bz(1 - 3z + z^2 + z^3)$$

$$- \tau y(z-y)(z + z^2 - 2y))$$

$$\times \left[\frac{1}{\mathcal{D}_2} + \frac{1}{\mathcal{D}_2(\alpha)} - \frac{2}{\mathcal{D}_2(\alpha) + y(1-\alpha)}\right]$$

$$+ \frac{\tau}{4}\int_0^1 dz \int_0^z dy (b(9 - 13z + 4z^2) - 2\tau y(z-y))$$

$$\times [\ln \mathcal{D}_2 + \ln \mathcal{D}_2(\alpha) - 2\ln(\mathcal{D}_2(\alpha) + y(1-\alpha))]$$

$$+ \frac{3\tau}{4}\int_0^1 dz \int_0^z dy [\mathcal{D}_2 \ln \mathcal{D}_2 + \mathcal{D}_2(\alpha) \ln \mathcal{D}_2(\alpha)$$

$$- 2(\mathcal{D}_2(\alpha) + y(1-\alpha))\ln(\mathcal{D}_2(\alpha) + y(1-\alpha))],$$

$$\bar{f}_Q^{(5)+(6)}(q^2) = \int_0^1 dz \int_0^z dy (a - bz)\frac{1}{\mathcal{D}_2(\alpha) + y(1-\alpha)}$$

$$- b\int_0^1 dz \int_0^z dy(1-z)((1-z)(a-bz) - \tau y(z-y))$$

$$\times \left[\frac{1}{\mathcal{D}_2(\alpha) + y(1-\alpha)} - \frac{1}{\mathcal{D}_2(\alpha)}\right] \quad (3.5)$$

$$- \frac{1}{2}\int_0^1 dz \int_0^z dy (a + 5b - 6bz)$$

$$\times [\ln(\mathcal{D}_2(\alpha) + y(1-\alpha)) - \ln \mathcal{D}_2(\alpha)],$$

where

$$a = \left(\frac{m_l}{M_W}\right)^2, \quad b = \left(\frac{m_\nu}{M_W}\right)^2,$$

$$\mathcal{D}_1(\alpha) = \alpha + (a-\alpha)z - bz(1-z) + \tau y(z-y),$$

$$\mathcal{D}_1 = \mathcal{D}_1(\alpha = 1) = 1 + (a-1)z - bz(1-z)$$
$$+ \tau y(z-y),$$

$$\mathcal{D}_2(\alpha) = a + (\alpha - a)z - bz(1-z) + \tau y(z-y),$$

$$\mathcal{D}_2 = \mathcal{D}_2(\alpha = 1)$$
$$= a + (1-a)z - bz(1-z) + \tau z(z-y).$$

Note that the values of the mass parameters of a charge lepton ($a$) and a neutrino ($b$) are exactly taken into account in expressions (3.1)–(3.5). The value of the gauge parameter $\alpha$ is arbitrary. The calculations were made for an arbitrary value of $q^2$.

The contributions from the $\gamma$–$Z$ diagrams shown in Fig. 4 to the charge form factor of a neutrino can be obtained on the basis of expansion (2.32) and have the form

$$f_Q^{(j)}(q^2) = \frac{g}{4\cos\theta_W}\frac{A^{(j)}(\alpha, q^2) + B^{(j)}(\alpha, q^2)}{q^2 - M_Z^2}, \quad (3.6)$$

$$j = 7, \ldots, 14.$$

Using the explicit form of functions $A^{(j)}(\alpha, q^2)$ (formulas (2.17)–(2.23)) and $B^{(j)}(\alpha, q^2)$ (formulas (2.24)–(2.30)) and employing relation (3.6), we can derive the expressions for the contributions from the $\gamma$–$Z$ diagrams for an arbitrary values of gauge parameter $\alpha$ and for $q^2 \neq 0$. However, these formulas are cumbersome and are not given here.

## 4. MAGNETIC FORM FACTOR OF A NEUTRINO

Relation (2.2) representing the general expansion of the neutrino electromagnetic vertex function $\Lambda_\mu(q)$ implies that the neutrino magnetic form factor $f_M(q^2)$ is the coefficient in the term proportional to $i\sigma_{\mu\nu}q^\nu$. In this section, we give the exact expression for $f_M(q^2)$ taking into account the dependence on two mass parameters $a$ and $b$ as well as on parameter $\alpha$ fixing the gauge.

Note that the Feynman diagrams depicted in Fig. 4 make zero contribution to the neutrino magnetic form factor. Thus, the exact expression for the neutrino magnetic form factor has the form

$$f_M(q^2) = \frac{eG_F}{4\pi^2\sqrt{2}}m_\nu \sum_{i=1}^{6} \bar{f}_M^{(i)}(q^2),$$

where coefficients $\bar{f}_M^{(i)}(q^2)$ are the contributions from the corresponding diagrams shown in Fig. 2 to the neutrino magnetic form factor. For this coefficients, we have the following relations:

$$\bar{f}_M^{(1)}(q^2) = \int_0^1 dz \int_0^z dy (2 - 3z + z^2)\frac{1}{D_1} - \frac{1}{2}\int_0^1 dz$$

$$\times \int_0^z dy(az^2 - bz^2(1-z) - ty(z-y)(2-z))$$

$$\times \left[\frac{1}{D_1(\alpha)} - \frac{1}{D_1}\right]$$

$$+ \frac{1}{2}\int_0^1 dz \int_0^z dy(2 - 3z)[\ln D_1(\alpha) - \ln D_1], \quad (4.1)$$

$$\bar{f}_M^{(2)}(q^2)$$
$$= \frac{1}{2}\int_0^1 dz \int_0^z dy z(a + az - b(1-z))\frac{1}{D_1(\alpha)}, \quad (4.2)$$

$$\bar{f}_M^{(3)}(q^2) = \frac{1}{2}\int_0^1 dz$$
$$\times \int_0^z dy(2a - 3az + az^2 - bz(1-z))\frac{1}{D_2(\alpha)}, \quad (4.3)$$

$$\bar{f}_M^{(4)}(q^2) = \frac{1}{2}\int_0^1 dz \int_0^z dy(1 + 2z)\frac{1}{D_2}$$

$$+ \frac{1}{2}\int_0^1 dz \int_0^z dy(b(1-z)^2(z(1-z) - 2y)$$

$$- ty(z-y)(2y - 3z + z^2) - 2ty)$$

$$\times \left[\frac{1}{D_2(\alpha) + y(1-\alpha)} - \frac{1}{D_2}\right]$$

$$+ \frac{1}{2}\int_0^1 dz \int_0^z dy(-2 + 9z - 4z^2 - 6y)$$

$$\times [\ln(D_2(\alpha) + y(1-\alpha)) - \ln D_2] - \frac{t}{4}\int_0^1 dz \quad (4.4)$$

$$\times \int_0^z dy(bz(1 - 3z + z^2 + z^3) - ty(z-y)(2 - z - z^2))$$

$$\times \left[\frac{1}{D_2} + \frac{1}{D_2(\alpha)} - \frac{2}{D_2(\alpha) + y(1-\alpha)}\right]$$

$$+ \frac{t}{8}\int_0^1 dz \int_0^z dy(8 - 13z + 3z^2)$$

$$\times [\ln D_2 + \ln D_2(\alpha) - 2\ln(D_2(\alpha) + y(1-\alpha))],$$

$$\bar{f}_M^{(5)+(6)}(q^2) = \int_0^1 dz \int_0^z dy y\frac{1}{D_2(\alpha) + y(1-\alpha)}$$

$$+ \frac{1}{2}\int_0^1 dz \int_0^z dy((a - bz)(1-z)^2 + ty(z-y)(1-z))$$

$$\times \left[\frac{1}{D_2(\alpha) + y(1-\alpha)} - \frac{1}{D_2(\alpha)}\right] + \frac{1}{2}\int_0^1 dz \quad (4.5)$$

$$\times \int_0^z dy(2 - 3z)[\ln(D_2(\alpha) + y(1-\alpha)) - \ln D_2(\alpha)],$$

here,

$$D_1(\alpha) = \alpha + (a - \alpha)z - bz(1-z) + ty(z-y),$$
$$D_1 = D_1(\alpha = 1)$$
$$= 1 + (a-1)z - bz(1-z) + ty(z-y),$$
$$D_2(\alpha) = a + (\alpha - a)z - bz(1-z) + ty(z-y),$$
$$D_2 = D_2(\alpha = 1)$$
$$= a + (1-a)z - bz(1-z) + ty(z-y),$$
$$t = -q^2/M_W^2.$$

Let us analyze the asymptotic behavior of the integrals contained in the contributions from triangular diagrams to $f_M(q^2)$ for large positive values of $t$. For example, let us consider the following integral for $t \longrightarrow \infty$:

$$J(t) = t\int_0^z dy \frac{y}{D_2(\alpha)} = \int_0^z dy \frac{y}{(y - y_2)(y_1 - y)}, \quad (4.6)$$

here,

$$y_1 = z + \frac{D}{zt} + \ldots, \quad y_2 = -\frac{D_\alpha}{zt} + \ldots \quad (4.7)$$

Carrying out elementary integration, we obtain

$$J(t) \longrightarrow \ln t - \ln D. \quad (4.8)$$

In formulas (4.6)–(4.8), we omitted the terms proportional to $1/t$ and $(\ln t)/t$, which are infinitely small for large positive values of $t$. The remaining integrals can be estimated similarly. Finally, we find that

$$\bar{f}_M(t) = \sum_{i=1}^{6} \bar{f}_M^{(i)}(t) \longrightarrow 0 \text{ as } t \longrightarrow \infty.$$

The asymptotic behavior of the magnetic form factor for large negative values of $q^2$ described here is in accordance with the Weinberg general theorem [24]. However, the case of the magnetic form factor of a massive neutrino has never been analyzed before. It should be noted that, while deriving relations (4.6)–(4.8), we assumed that $\alpha < \infty$. Thus, the obtained result ($\bar{f}_M(t) \longrightarrow 0$ as $t \longrightarrow 0$) is valid for any gauge except the unitary one. The value of $\bar{f}_M(t \longrightarrow \infty)$ may differ from zero if we first set $\alpha = \infty$ and then proceed to the limit $t \longrightarrow +\infty$. The behavior of magnetic form factors in the framework of the Weinberg–Salam model in the unitary gauge is analyzed, for example, in [14]. Using explicit expressions for the magnetic form factor of a massive neutrino for an arbitrary value of gauge parameter $\alpha$, Fig. 5 shows the behavior of function $\bar{f}_M(t)$ in various gauges for a wide range of $t$ values: $0 \le t \le 5 \times 10^{-4}$. It can be seen that the magnetic form factor becomes independent of the choice of the gauge at $t = 0$, which corresponds to a photon on the mass surface. The value of $f_M(t = 0)$ is equal to the neutrino magnetic moment. The fact that the magnetic moment of the massive neutrino is independent of the choice of the gauge

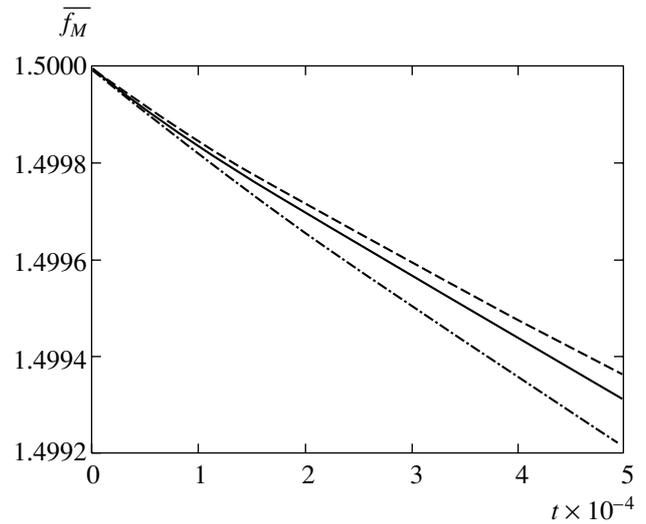

**Fig. 5.** Magnetic form factor of a massive neutrino as a function of $t$ for various values of the gauge parameter. The dashed curve corresponds to $\alpha = 100$, the solid curve to the 't Hoft–Feynman gauge ($\alpha = 1$), and the dot-and-dash curve, to $\alpha = 0.1$.

was proved by direct calculation in our previous publication [13].

## 5. ANAPOLE FORM FACTOR OF A NEUTRINO

In this section, we consider the anapole form factor of a massive neutrino. Using the results obtained in Section 2 for various contributions to the neutrino vertex function $\Lambda_\mu(q)$, we single out in formulas (2.3)–(2.15) the coefficients proportional to $(q^2\gamma_\mu - \slashed{q}q_\mu)\gamma_5$, which are, in accordance with the expansion in relation (2.2), the corresponding contributions to the anapole form factor $f_A(q^2)$. It should be noted that, while separating similar terms in the contributions from each of the diagrams depicted in Figs. 2 and 4, we inevitably obtain additional terms proportional to matrix $\gamma_\mu\gamma_5$. Consequently, it is necessary to make sure that the corresponding form factor has zero value even at $q^2 \ne 0$. This problem was also touched upon in Section 2.

We will first consider the contributions from single-loop triangular diagrams (see Fig. 2) to the neutrino anapole form factor. Integrating with respect to momenta of virtual particles (see the monograph [23]), we obtain exact expression for the contributions of these diagrams to the anapole form factor of a massive neutrino in terms of definite integrals,

$$f_A(q^2) = \frac{eG_F}{4\pi^2\sqrt{2}} \sum_{i=1}^{6} \bar{f}_A^{(i)}(q^2),$$

where

$$\bar{f}_A^{(1)}(q^2) = -\frac{1}{2}\int_0^1 dz \int_0^z dy((2-z-z^2) + 4y(z-y))\frac{1}{\mathfrak{D}_1}$$

$$-\frac{1}{4}\int_0^1 dz \int_0^z dy[(a+b(1-z))z^2$$

$$-4y(z-y)(a-b) - 2\tau y(z-y)(2-z)]\left[\frac{1}{\mathfrak{D}_1(\alpha)} - \frac{1}{\mathfrak{D}_1}\right]$$

$$-\frac{1}{4}\int_0^1 dz \int_0^z dy(2-3z)[\ln\mathfrak{D}_1(\alpha) - \ln\mathfrak{D}_1], \quad (5.1)$$

$$\bar{f}_A^{(2)}(q^2) = -\frac{a-b}{4}$$
$$\times \int_0^1 dz \int_0^z dy(z(1-z) + 4y(z-y))\frac{1}{\mathfrak{D}_1(\alpha)}, \quad (5.2)$$

$$\bar{f}_A^{(3)}(q^2) = \frac{a-b}{4}\int_0^1 dz \int_0^z dy(z-2y)^2 \frac{1}{\mathfrak{D}_2(\alpha)}, \quad (5.3)$$

$$\bar{f}_A^{(4)}(q^2) = -\frac{1}{4}\int_0^1 dz \int_0^z dy(z(3-2z) + 8y(z-y))\frac{1}{\mathfrak{D}_2}$$

$$-\frac{1}{4}\int_0^1 dz \int_0^z dy[bz(1-z^2)(3-z)$$

$$-4b(1-z)(z-y(1+z))(z-y)$$

$$-\tau(4y^3 - 4y^2z - 2y + yz^2 - 4y^2 + yz)(z-y)]$$

$$\times \left[\frac{1}{\mathfrak{D}_2(\alpha) + y(1-\alpha)} - \frac{1}{\mathfrak{D}_2}\right]$$

$$+\frac{1}{4}\int_0^1 dz \int_0^z dy(6-3z-4z^2+12y+16y(z-y)) \quad (5.4)$$

$$\times [\ln(\mathfrak{D}_2(\alpha) + y(1-\alpha)) - \ln\mathfrak{D}_2]$$

$$-\frac{1}{4}\int_0^1 dz \int_0^z dy[b(1-z^2) + \tau(z+2y(z-y))]$$

$$\times [\ln\mathfrak{D}_2 + \ln\mathfrak{D}_2(\alpha) - 2\ln(\mathfrak{D}_2(\alpha) + y(1-\alpha))]$$

$$+\frac{3}{4}\int_0^1 dz \int_0^z dy[\mathfrak{D}_2 \ln\mathfrak{D}_2 + \mathfrak{D}_2(\alpha)\ln\mathfrak{D}_2(\alpha)$$

$$-2(\mathfrak{D}_2(\alpha) + y(1-\alpha))\ln(\mathfrak{D}_2(\alpha) + y(1-\alpha))],$$

$$\bar{f}_A^{(5)+(6)}(q^2) = \frac{1}{2}\int_0^1 dz \int_0^z dy\, y\frac{1}{\mathfrak{D}_2(\alpha) + y(1-\alpha)}$$

$$+\frac{1}{4}\int_0^1 dz \int_0^z dy[a((1+z)^2 - 4y(1+z-y))$$

$$-b(z(1+z)^2 - 2y((1+z)^2 - 2y))$$

$$+\tau y(z-y)(1+z-2y)]\left[\frac{1}{\mathfrak{D}_2(\alpha) + y(1-\alpha)} - \frac{1}{\mathfrak{D}_2(\alpha)}\right] \quad (5.5)$$

$$-\frac{1}{4}\int_0^1 dz \int_0^z dy(2+3z-6y)$$

$$\times [\ln(\mathfrak{D}_2(\alpha) + y(1-\alpha)) - \ln\mathfrak{D}_2(\alpha)].$$

Note that the values of the mass parameters of a charged lepton ($a$) and a neutrino ($b$) are exactly taken into account in expressions (5.1)–(5.5). The value of gauge parameter $\alpha$ is arbitrary. All calculations were made for an arbitrary value of $q^2$.

The contributions from the $\gamma$–$Z$ diagrams shown in Fig. 4 to the neutrino anapole form factor can be derived using expansion (2.32) and have the form

$$f_A^{(j)}(q^2) = \frac{g}{4\cos\theta_W}\frac{1}{q^2 - M_Z^2}$$

$$\times \left\{\frac{A^{(j)}(\alpha, q^2)}{q^2} + (1-\alpha_Z)\frac{B^{(j)}(\alpha, q^2)}{q^2 - \alpha_Z M_Z^2}\right\}, \quad (5.6)$$

$$j = 7, ..., 14.$$

Using the explicit form of functions $A^{(j)}(\alpha, q^2)$ (formulas (2.17)–(2.23)) and $B^{(j)}(\alpha, q^2)$ (formulas (2.24)–(2.30)) and employing relation (5.6), we can derive expressions for the contributions from the $\gamma$–$Z$ diagrams for arbitrary values of gauge parameters $\alpha$ and $\alpha_Z$ and for $q^2 \neq 0$. However, these formulas are cumbersome and will not be given here.

*Anapole Moment*

Let us consider the anapole moment of a massive neutrino. We have obtained the contributions from the triangular diagrams (formulas (5.1)–(5.5)) as well as from the $\gamma$–$Z$ diagrams (5.6) to the neutrino anapole form factor for an arbitrary value of $q^2$. Since the anapole moment is a static electromagnetic parameter of the neutrino, the value of $q^2$ should be set equal to zero in the formulas under consideration.

In the case of a zero-mass neutrino, the value of the anapole moment is connected with the charge radius through the relation (see, for example, [21])

$$a_\nu = \frac{1}{6}\langle r_\nu^2 \rangle.$$

However, in the case of a massive particle, this simple relationship is violated for the reasons described in Section 2.

In the expression for the anapole moments, the contributions come both from the triangular diagrams depicted in Fig. 2 and from the $\gamma$–Z diagrams shown in Fig. 4. Thus, the complete expression for the anapole moment has the form

$$a_\nu = \frac{eG_F}{4\pi^2\sqrt{2}}\left\{\sum_{i=1}^{6}\bar{a}^{(i)}(a, b, \alpha) + \sum_{j=7}^{14}\bar{a}^{(j)}(a, \alpha)\right\}.$$

The contributions $\bar{a}^{(i)}(a, b, \alpha)$ from the triangular diagrams can be written in the form

$$\bar{a}^{(1)}(a, b, \alpha) = -\frac{1}{6}\int_0^1 dz(2 + 3z - 6z^2 + z^3)\frac{1}{D}$$

$$-\frac{1}{12}\int_0^1 dz(1-z)^3(a+2b+3bz)\left[\frac{1}{D_\alpha} - \frac{1}{D}\right] \quad (5.7)$$

$$+\frac{1}{4}\int_0^1 dz(1-z)(1-4z+3z^2)[\ln D_\alpha - \ln D],$$

$$\bar{a}^{(2)}(a, b, \alpha) = -\frac{a-b}{12}\int_0^1 dz(2-3z+z^3)\frac{1}{D_\alpha}, \quad (5.8)$$

$$\bar{a}^{(3)}(a, b, \alpha) = \frac{a-b}{12}\int_0^1 dz z^3 \frac{1}{D_\alpha}, \quad (5.9)$$

$$\bar{a}^{(4)}(a, b, \alpha) = -\frac{1}{12}\int_0^1 dz z^2(9-2z)\frac{1}{D} - \frac{b}{4}\int_0^1 dz$$

$$\times \int_0^z dy[z(1-z^2)(3-z) - 4(1-z)(z-y(1+z))(z-y)]$$

$$\times \left[\frac{1}{D_\alpha + y(1-\alpha)} - \frac{1}{D}\right]$$

$$+\frac{1}{4}\int_0^1 dz \int_0^z dy(6 - 3z - 4z^2 + 12y + 16y(z-y))$$
(5.10)
$$\times [\ln(D_\alpha + y(1-\alpha)) - \ln D] - \frac{b}{4}\int_0^1 dz$$

$$\times \int_0^z dy(1-z^2)[\ln D + \ln D_\alpha - 2\ln(D_\alpha + y(1-\alpha))]$$

$$+\frac{3}{4}\int_0^1 dz \int_0^z dy[D\ln D + D_\alpha \ln D_\alpha$$

$$- 2(D_\alpha + y(1-\alpha))\ln(D_\alpha + y(1-\alpha))],$$

$$\bar{a}^{(5)+(6)}(a, b, \alpha) = \frac{1}{2}\int_0^1 dz \int_0^z dy y \frac{1}{D_\alpha + y(1-\alpha)}$$

$$+\frac{1}{4}\int_0^1 dz \int_0^z dy[a((1+z)^2 - 4y(1+z-y))$$

$$- b(x(1+z)^2 - 2y((1+z)^2 - 2y))] \quad (5.11)$$

$$\times \left[\frac{1}{D_\alpha + y(1-\alpha)} - \frac{1}{D_\alpha}\right] - \frac{1}{4}\int_0^1 dz \int_0^z dy(2+3z-6y)$$

$$\times [\ln(D_\alpha + u(1-\alpha)) - \ln D_\alpha].$$

It should be noted that, as in the case of a zero-mass neutrino, the contributions from triangular diagrams to the anapole moment are finite.

To obtain the contributions from the $\gamma$–Z diagrams to the anapole moment, it is convenient to use formula (2.32), setting $q^2 = 0$ in this case. Thus, the expressions for

$$a^{(j)}(a, \alpha) = \frac{eG_F}{4\pi^2\sqrt{2}}\bar{a}^{(j)}(a, \alpha)$$

assume the form

$$a^{(j)}(a, \alpha) = \frac{1}{M_Z^2}Q^{(j)} - \frac{g}{4M_Z^2\cos\theta_W}\frac{A^{(j)}(q^2)}{q^2}\bigg|_{q^2 \to 0}, \quad (5.12)$$

$$j = 1, ..., 14,$$

where the contributions to the neutrino electric charge $Q^{(j)}$ can be derived from expressions (2.24)–(2.30) and (2.32) (see also our previous publication [13]). While deriving relation (5.12), we assumed that $\alpha_Z = \infty$. Using formula (5.12), we obtain the following expressions for $\bar{a}^{(i)}(a, \alpha)$:

$$\bar{a}^{(7)}(a, \alpha) = \cos^2\theta_W \left\{ \cos^2\theta_W \left[ \omega\left(3 + \frac{3}{3}\alpha(1+\alpha)\right) \right. \right.$$

$$\left. -1 - \frac{5\alpha}{8} - \frac{5\alpha^2}{8} - \frac{3\alpha^3}{4}\frac{\ln\alpha}{1-\alpha} \right]$$

$$-\frac{1}{2}\left[ \omega\left(-\frac{14}{3} + \alpha\right) - \frac{1}{18(1-\alpha)^3} \right. \quad (5.13)$$

$$\times (11 - 54\alpha + 54\alpha^2 - 2\alpha^3 - 9\alpha^4$$

$$\left. \left. - 18\alpha^2\ln\alpha - 12\alpha^3\ln\alpha + 18\alpha^4\ln\alpha) \right] \right\},$$

$$\bar{a}^{(8)}(a, \alpha) = \sin^2\theta_W \left\{ \cos^2\theta_W \left[ \omega\frac{3+\alpha}{4} - \frac{5+\alpha}{8} \right. \right.$$

$$\left. -\frac{\alpha}{2}\left(1 + \frac{\alpha}{2}\right)\frac{\ln\alpha}{1-\alpha} \right] - \frac{1}{18(1-\alpha)^3} \quad (5.14)$$

$$\left. \times (11 - 18\alpha + 9\alpha^2 - 2\alpha^3 + 8\ln\alpha) \right\},$$

$$\bar{a}^{(9)}(\alpha, \alpha) = (\cos^2\theta_W - \sin^2\theta_W)\cos^2\theta_W$$
$$\times \frac{1}{2}\{-\omega\alpha + \alpha - \alpha\ln\alpha\}, \quad (5.15)$$

$$\bar{a}^{(10)}(a, \alpha) = \cos^4\theta_W$$
$$\times \left\{ -\frac{3}{4}\omega(3+\alpha^2) + \frac{3}{8} + \frac{5\alpha^2}{8} - \frac{3}{4}\alpha^2\ln\alpha \right\}, \quad (5.16)$$

$$\bar{a}^{(11)+(12)}(a, \alpha) = \cos^2\theta_W$$
$$\times \frac{1}{2}\left\{ \cos^2\theta_W(-\omega\alpha + \alpha) - \frac{1}{3}(\omega + \ln\alpha) \right\}, \quad (5.17)$$

$$\bar{a}^{(13)}(a, \alpha) = (\sin^2\theta_W - \cos^2\theta_W)$$
$$\times \frac{1}{2}\left\{ \cos^2\theta_W(-\omega\alpha + \alpha - \alpha\ln\alpha) + \frac{1}{6}(\omega + \ln\alpha) \right\}, \quad (5.18)$$

$$\bar{a}^{(14)}(a, \alpha) = -\left\{ \omega\left(-1 - \frac{28}{9}\sin^2\theta_W\right) \right.$$

$$\left. + \frac{1}{3}\sum_f Q_f\left(\pm\frac{1}{2} - 2Q_f\sin^2\theta_W\right)\ln\left(\frac{m_f}{M}\right)^2 \right\}. \quad (5.19)$$

Expressions (5.13)–(5.19) are diverging; consequently, the final form of these formulas depends on the method for regularizing Feynman integrals (2.9)–(2.15). This circumstance can be used to explain a certain difference between relations (5.13)–(5.19) and the corresponding contributions to the charge radius, which were derived in [16].

The direct calculation performed here gives diverging expressions for the anapole moment of a massive neutrino. It should be recalled in this connection that the corresponding corrections to the electromagnetic vertex function can be treated from the standpoint of radiative corrections to the expression for a physical process (e.g., neutrino scattering by a charged lepton). It is quite obvious that the scattering cross section, which is a measurable quantity, must be finite and independent of the choice of the gauge. It is precisely this approach, which was developed in [17, 21, 25, 26] for the case of a zero-mass neutrino, that formed the basis of the definition of the electroweak anapole moment and the electroweak charge radius. Note that, in addition to the Feynman diagrams considered here, the corrections to the electromagnetic vertex function of a charged lepton should also be taken into account in studying the radiative corrections to scattering. Moreover, the so-called "box" diagram, in which the neutrino and the lepton exchange two virtual bosons in the course of the interaction, will also affect the scattering process. Detailed calculations of the corresponding diagrams for the case of zero-mass neutrino are given in the recent publications [25, 26]. It would be interesting to demonstrate that, for a massive neutrino also, the diverging terms depending on the gauge parameters will cancel out. However, in the case of a massive neutrino, this problem is not trivial and requires an additional independent analysis.

## 6. ELECTROMAGNETIC CHARACTERISTICS OF A NEUTRINO IN THE CASE OF MIXING BETWEEN DIFFERENT GENERATIONS

It was noted in Section 1 that contemporary experimental data speak in favor of mixing existing between different generations of neutrinos. In the study of the electromagnetic properties of neutrinos, this is manifested in the existence of transient electromagnetic dipole moment (and also transient form factors) of a neutrino. Indeed, formula (2.1) in the present case assumes the form

$$\langle \nu(p') | J_\mu^{EM} | \nu(p) \rangle = \sum_{\alpha\beta} \bar{u}_\beta(p') \Lambda_\mu^{\beta\alpha}(q) u_\alpha(p),$$

where summation is carried out over neutrino types ($\nu_e$, $\nu_\mu$ and $\nu_\tau$). In analogy with expansion (2.2), we conclude that the most general expression for vertex function $\Lambda_\mu^{\beta\alpha}(q)$ is

$$\begin{aligned}\Lambda_\mu^{\beta\alpha}(q) &= f_Q^{\beta\alpha}\gamma_\mu + f_M^{\beta\alpha} i\sigma_{\mu\nu} q^\nu - f_E^{\beta\alpha}\sigma_{\mu\nu} q^\nu \gamma_5 \\ &+ f_A^{\beta\alpha}(q^2 \gamma_\mu - q_\mu \slashed{q})\gamma_5,\end{aligned} \quad (6.1)$$

where the quantities $f_I^{\beta\alpha}$ ($I = Q, M, E, A$) have the meaning of transient form factors.

By way of an example, we analyze the slightly extended standard model with mixing between different generations of charged leptons and neutrinos. The interaction Lagrangian and the Feynman rules are given in [22]. First, we consider the situation when $m_{\nu_e} = m_{\nu_\mu} = m_\nu$. Such a choice of parameters corresponds to three generations of neutrinos degenerate in mass. In this case, we can prove that expressions for $f_I^{\beta\alpha}$ have the form

$$f_I^{\beta\alpha} = \sum_{l = e,\mu,\tau} U_{\beta l} \overset{*}{U}_{\alpha l} f_I^{(l)}, \quad (6.2)$$

where $U_{\beta l}$ is the unitary matrix describing mixing between different generations of leptons and neutrinos (see review [22]) and $f_I^{(l)} = f_I(a_l, b, q^2, ...)$ are the expressions for form factors derived in Sections 3–5 of this paper; $a_l = (m_l/M_W)^2$.

In actual practice, the case of almost degenerate neutrino masses is not ruled out by the available experimental data (see [27, 28]). Considering the case of mass-degenerate neutrinos, we in fact introduce two additional (apart from parameters $a = (m_l/M_W)^2$ and $b = (m_\nu/M_W)^2$ used above) mass parameters $c_i = (\Delta m_i/M_W)^2$, where $\Delta m_i$ are two independent values of the mass differences between three flavored neutrinos, and carry out the expansion in these parameters in subsequent calculations, assuming that $c \ll a, b$.

If we set $q^2 = 0$ in formula (6.2), this will lead to expressions for transient charges, which determine the static electromagnetic properties of three neutrino types. It is interesting to note that the structure of relation (6.2) and the results obtained in out previous publications [13] imply that the values of neutrino transient charges $Q_{\beta\alpha} = f_Q^{\beta\alpha}(q^2 = 0)$ are identically equal to zero for an arbitrary value of gauge parameter $\alpha$.

In the case of nondegenerate neutrino masses, apart from the three form factors considered in this section, relation (6.1) also acquires the transient electric dipole form factor that is identically equal to zero for neutrinos with degenerate masses. For this reason, we will give the expression only for this form factor. The simplest expression for the electric dipole form factor is obtained in the 't Hoft–Feynman gauge. Note that the contributions to this form factor comes only from triangular diagrams. On the basis of expressions (2.3)–(2.7) (substituting $m_\beta$ for $m_\nu$ in the extreme left parentheses and $m_\alpha$ for $m_\nu$ in the extreme right parentheses in formulas (2.4) and (2.5), where $m_{\beta,\alpha}$ are the flavor masses of the final and the initial neutrino states), averaging over the initial and final neutrino states, we find that

$$f_E^{\beta\alpha}(q^2) = \frac{eG_F}{4\pi^2 \sqrt{2}} \sum_{i=1}^{6} \left( \sum_{l = e,\mu,\tau} U_{\beta l} U_{\alpha l} \bar{E}_{\beta\alpha}^{(i)}(q^2, a_l) \right).$$

Here, the contributions from each Feynman diagram have the form

$$\begin{aligned}\bar{E}_{\beta\alpha}^{(1)} = &-\frac{i}{2}\int_0^1 dx \int_0^{1-x} dy \frac{1}{D_{\beta\alpha}^{(1)}} \\ &\times [(m_\beta - m_\alpha)(2 - 3(x+y) + (x+y)^2) \\ &+ (m_\beta + m_\alpha)(x - y - (x^2 - y^2))],\end{aligned} \quad (6.3)$$

$$\begin{aligned}\bar{E}_{\beta\alpha}^{(2)} = &-\frac{i}{4}\int_0^1 dx \int_0^{1-x} dy \frac{1}{D_{\beta\alpha}^{(1)}} \\ &\times \{a_l[(m_\beta - m_\alpha)(x + y + (x+y)^2) \\ &+ (m_\beta + m_\alpha)(x - y - (x^2 - y^2))] \\ &+ \sqrt{b_\alpha b_\beta}[(m_\beta - m_\alpha)((x+y) - (x+y)^2) \\ &- (m_\beta + m_\alpha)((x-y) - (x^2 - y^2))]\},\end{aligned} \quad (6.4)$$

$$\begin{aligned}\bar{E}_{\beta\alpha}^{(3)} = &\frac{i}{4}\int_0^1 dx \int_0^{1-x} dy \frac{1}{D_{\beta\alpha}^{(2)}}(x+y-1) \\ &\times [(m_\beta - m_\alpha)(2 a_l + (x+y)(\sqrt{b_\alpha b_\beta} - a_l)) \\ &- (m_\beta + m_\alpha)(x - y)(\sqrt{b_\alpha b_\beta} - a_l)],\end{aligned} \quad (6.5)$$

$$\begin{aligned}\bar{E}_{\beta\alpha}^{(4)} = &-\frac{i}{4}\int_0^1 dx \int_0^{1-x} dy \frac{1}{D_{\beta\alpha}^{(2)}} \\ &\times [(m_\beta - m_\alpha)(x + y + 2(x+y)^2) \\ &+ (m_\beta + m_\alpha)(x - y - 2(x^2 - y^2))],\end{aligned} \quad (6.6)$$

$$\bar{E}_{\beta\alpha}^{(5)+(6)} = \frac{i}{4}\int_0^1 dx \int_0^{1-x} dy \frac{1}{D_{\beta\alpha}^{(2)}} \quad (6.7)$$
$$\times [(m_\beta - m_\alpha)(x+y) + (m_\beta + m_\alpha)(x-y)]$$

and

$$D_{\beta\alpha}^{(1)} = 1 - (a_l - 1)(x+y) - \frac{b_\beta + b_\alpha}{2}(x+y)(1-x-y)$$
$$- \tau xy + \frac{b_\beta + b_\alpha}{2}(x - y - (x^2 - y^2)),$$

$$D_{\beta\alpha}^{(2)} = a_l - (1 - a_l)(x+y)$$
$$- \frac{b_\beta + b_\alpha}{2}(x+y)(1-x-y)$$
$$- \tau xy + \frac{b_\beta - b_\alpha}{2}(x - y - (x^2 - y^2)),$$

$b_\alpha = (m_\alpha/M_W)^2$. We have derived formulas (6.3)–(6.7) using the fact that the relation

$$\bar{u}_\beta(p')(p'_\mu + p_\mu)\gamma_5 u_\alpha(p)$$
$$= \bar{u}_\beta(p')[(m_\beta - m_\alpha)\gamma_\mu\gamma_5 - i\sigma_{\mu\nu}\gamma_5 q^\nu]u_\alpha(p),$$

holds for a neutrino in the mass shell also.

It follows from formulas (6.3)–(6.7) that, in the case of degenerate neutrino masses (i.e., for $m_\beta = m_\alpha$), the contributions from each Feynman diagram to the electric dipole form factor are identically equal to zero.

On the basis of relations (6.3)–(6.7), we can derive the expressions for transient electric dipole moments of neutrinos. To this end, we must set $\tau = 0$ in the formulas under investigation. The simplest expressions for the electric dipole moments are obtained for light neutrinos ($b_\alpha \ll 1$). Expanding the integrands in formulas (6.3)–(6.7) in parameter $b_\alpha$ and integrating with respect to Feynman parameters $x$ and $y$, we obtain

$$\sum_{i=1}^{6} \bar{E}_{\beta\alpha}^{(i)}(q^2 = 0, a_l) = -\frac{i}{24(1-a_l)^4}(m_\beta - m_\alpha)$$
$$\times (12 - 52a_l + 81a_l^2 - 48a_l^3 + 7a_l^4 \quad (6.8)$$
$$- 6a_l^2 \ln a_l + 12a_l^3 \ln a_l).$$

Proceeding from the fact that charged leptons must be much lighter than a $W$ boson (i.e., $a_l \ll 1$), we obtain from relation (6.8) the final expression for transient electric dipole moments of the neutrino in the form

$$d_{\beta\alpha} = f_E^{\beta\alpha}(q^2 = 0) = \frac{eG_F}{8\pi^2\sqrt{2}} i(m_\alpha - m_\beta). \quad (6.9)$$

Note that this relation does not contain the dependence on the mixing angle since we disregard the masses of charged leptons. It is obvious from formula (6.9) that, in the slightly extended standard model with mixing between different generations of charged leptons and neutrinos, the neutrino electric dipole moments that are diagonal in flavors are equal to zero, while nondiagonal elements are proportional to the difference in the neutrino flavor masses.

## 7. CONCLUSIONS

We have investigated the electromagnetic vertex form factors of a massive Dirac neutrino in the framework of the slightly extended standard model supplemented with a SU(2)-singlet right neutrino. In all calculations, we have exactly taken into account the masses of a charged lepton and a neutrino. Calculations were made in an arbitrary $R_\xi$ gauge, which makes it possible to analyze the dependence of the obtained results on the gauge parameters of both $W$ and $Z$ bosons. It was found in Section 2 that, for a certain choice of gauge parameters, all electromagnetic form factors of the neutrino become finite (i.e., contain no ultraviolet divergence). This statement has been proved in the one-loop approximation. An analogous property of the electromagnetic vertex function can be formulated in the framework of quantum electrodynamics. For a certain choice of the photon gauge parameter ($\alpha_\gamma = 3$), the electron electromagnetic vertex function in the one-loop approximation does not contain infrared divergence. In Sections 3 and 4, the contributions from all Feynman diagrams to the charge and magnetic form factors, which exactly take into account the dependence on mass parameters $a$ and $b$ as well as on the gauge parameter $\alpha$, have been determined for the first time. The asymptotic behavior of the magnetic form factor of a massive neutrino is investigated for $q^2 \rightarrow -\infty$ and it is found that $f_M(q^2) \rightarrow 0$ in this case. The anapole form factor and the anapole moment of a massive neutrino are considered in Section 5 for an arbitrary value of gauge parameter $\alpha$. It is also found that, like in the case of a zero-mass particle, the anapole moment of a massive neutrino is a diverging quantity and depends of the choice of the gauge. The transient electromagnetic form factors of the neutrino are studied in Section 6 in the framework of the slightly extended standard model permitting mixing between different generations of charged leptons and neutrinos. Using the results obtained in Sections 3–5, we have obtained for the first time the explicit expressions for the transient charge, magnetic, and anapole form factors for neutrino states degenerate in flavor masses. It is shown that transient electric charges are identically equal to zero. For the case of neutrino states nondegenerate in masses, an exact expression for the transient electric dipole form factor is obtained in the 't Hoft–Feynman gauge. This form factor is identically equal to zero for mass-degenerate neutrinos. Moreover, the expression for the transient electric dipole moment was also found in this gauge. It was found that the transient electric dipole moment is proportional to the difference in the masses of the initial and final neutrino states.

*Translated by N. Wadhwa*

SPELL: 1. anapole, 2. hence,s